\documentclass[11pt]{article}
\usepackage{amssymb}
\usepackage{amsmath}
\usepackage[dvips]{graphicx}
\makeatletter \@addtoreset{equation}{section} \makeatother

\makeatletter
\renewcommand\section{\@startsection {section}{1}{\z@}%
                                   {-5.5ex \@plus -1ex \@minus -.2ex}
                                   {2.3ex \@plus.2ex}%
                                   {\normalfont\large\bfseries}}
\renewcommand\subsection{\@startsection{subsection}{2}{\z@}%
                                     {-3.25ex\@plus -1ex \@minus -.2ex}%
                                     {1.5ex \@plus .2ex}%
                                     {\normalfont\normalsize\bfseries}}
\renewcommand\thesection {\@arabic\c@section}
\renewcommand\thesubsection   {\thesection.\@arabic\c@subsection}
\renewcommand{\@seccntformat}[1]{%
\csname the#1\endcsname.\hspace{1.0em}}
\makeatother

\newcommand{\Tr}{\textrm{Tr}}

\newcommand{\ev}[1]{\big\langle #1 \big\rangle}
\newcommand{\dd}{\textrm{d}}
\newcommand{\LE}{\mathcal{L}_{\rm{E}}}
\newcommand{\SE}{S_{\rm{E}}}
\newcommand{\MSb}{\overline{\rm{MS}}}
\newcommand{\Lb}{\overline{\Lambda}}
\newcommand{\Nc}{N_{\rm c}}
\newcommand{\Nf}{N_{\rm f}}

\newcommand{\mub}{\bar\mu}
\newcommand{\dA}{d_{\rm A}}
\newcommand{\Tc}{T_{\rm c}}

\def\lsi{\raise0.3ex\hbox{$<$\kern-0.75em\raise-1.1ex\hbox{$\sim$}}}
\def\gsi{\raise0.3ex\hbox{$>$\kern-0.75em\raise-1.1ex\hbox{$\sim$}}}
\newcommand{\lsim}{\mathop{\lsi}}
\newcommand{\gsim}{\mathop{\gsi}}
\newcommand{\eq}[1]{Eq.~(\ref{#1})}

\newcommand{\aA}{\hat A_0}
\newcommand{\iz}{z_{\rm I}}
\newcommand{\rz}{z_{\rm R}}

\begin{document}

\begin{titlepage}
\begin{flushright}
HIP-2008-07/TH\\
NSF-KITP-08-10
\end{flushright}
\begin{centering}
\vfill{}
 
{\Large{\bf The diagonal and off-diagonal quark number 
    susceptibility of high temperature and finite density QCD}}

\vspace{0.8cm}

A. Hietanen$^{\rm a}$, 
K. Rummukainen$^{\rm b}$, 

\vspace{0.8cm}

{\em $^{\rm a}$%
Theoretical Physics Division, 
Department of Physics \\  
and Helsinki Institute of Physics, \\	
P.O.Box 64, FI-00014 University of Helsinki, Finland\\ 
}

{\em $^{\rm b}$%
Department of Physics, \\
University of Oulu P.O.Box 3000, FI-90014 Oulu,
Finland \\
}

\vspace*{0.3cm}
 
\noindent
\abstract{%
  We study the quark number susceptibility of the hot
  quark-gluon plasma at zero and non-zero quark number density,
  using lattice Monte Carlo simulations of an effective theory of QCD, 
  electrostatic QCD (EQCD).  Analytic continuation is used to
  obtain results at non-zero quark chemical potential $\mu$.
  We measure both flavor singlet
  (diagonal) and non-singlet (off-diagonal) quark number
  susceptibilities.   
  The diagonal susceptibility approaches the
  perturbative result above $\sim 20\Tc$, but below that temperature
  we observe significant deviations.  The results agree well 
  with 4d lattice data down to temperatures $\sim 2\Tc$.
  The off-diagonal
  susceptibility is more prone to statistical and systematic errors, but
  the results are consistent with perturbation theory already at
  $10\Tc$.

}
\vfill
\noindent
 


\vspace*{1cm}
 
\noindent

\vfill
\end{centering}
\end{titlepage}

%
\section{Introduction}

The quark (baryon) number susceptibility of hot QCD matter characterizes the
``softness'' of the equation of state.  It is directly related to the
event-by-event fluctuations observed in heavy ion collision
experiments \cite{Asakawa:2000wh}, probing the phase diagram and the
properties of the hot QCD plasma.  Thus, it is of significant interest to
calculate it theoretically as accurately as possible. Hence, 
several calculations of susceptibility have been published
using lattice simulations
\cite{gavai,Gavai:2002kq,karsch,Allton:2003vx,Bernard:2007nm,Maezawa:2007ew,Bluhm:2008sc} or perturbation
theory \cite{vuorinen02,Blaizot:2001vr,Ipp:2006ij}.

In this work we use lattice Monte Carlo simulations in order to
measure the diagonal (flavor singlet) and off-diagonal (non-singlet)
quark number susceptibilities at high temperatures and at non-zero
densities.  Instead of full 4-dimensional QCD, the theory
we study on the lattice is a dimensionally reduced effective theory of
the hot quark-gluon plasma phase of QCD, electrostatic QCD (EQCD)
\cite{ginsparg80,appelquist81,kajantie95,braaten95,Kajantie:1997tt}.
It is by now well established that EQCD can accurately describe many
properties of the hot QCD plasma, and it provides a very convenient
starting point for studying high-temperature QCD using perturbative
analysis \cite{gsixg,vuorinen03,gynther05} or non-perturbative lattice
simulations.

The validity of the effective theory approach is based on the fact
that at high enough temperatures the gauge coupling constant $g$
becomes small, giving rise to three relevant momentum scales
(neglecting quark masses): {\em hard} scale $p \sim\pi T$, corresponding
to non-zero Matsubara frequencies, {\em soft} electric scale $\sim gT$
and {\em supersoft} magnetic scale $\sim g^2T$.  EQCD is obtained by
(formally) integrating over the hard scales perturbatively, leaving an
effective theory for soft and supersoft scales.  All infrared
divergences inherent in finite temperature field theories are correctly
contained in the effective theory.  A crucial feature of EQCD is
that all of the fermionic modes are integrated over, leaving a 
purely bosonic theory.  

EQCD offers an interesting alternative to standard high-temperature
lattice simulations.  Above all, the theory is three-dimensional and
purely bosonic, making it much cheaper to simulate. The standard QCD
lattice simulations work well at temperatures up to 5--10 $\Tc$, but due
to the sheer cost of the simulations with light quarks it can be very
difficult to obtain accurate results.  In contrast the perturbative
analysis works at temperatures $T \gsim 10 \Tc$ (albeit with slow
convergence), but since the infrared singularities in the magnetic
sector cannot be treated perturbatively the accuracy is limited to
some order (depending on the observable) in the coupling constant expansion.  
The lattice simulations of EQCD fully include the effects of the
infrared singularities, thus offering a clear way to improve on the
perturbative results. 
While EQCD cannot describe the QCD phase transition, it has been
observed to be quantitatively accurate down to temperatures of
order 2--4 $T_c$, depending on the quantity of interest.
On the other hand, it is relatively easy to do EQCD simulations
at arbitrarily high temperatures, enabling one to quantify the
convergence to the perturbation theory and the role of the
infrared singularities.
Lattice simulations of EQCD have been used to
calculate QCD pressure at high temperature \cite{pressure,plaquette}, spatial
string tension \cite{spatialstring}, and spatial screening lengths
\cite{Kajantie:1997pd,hart00,hart02}.

In this paper we present the lattice calculations using EQCD to
measure the diagonal and off-diagonal quark number (baryon number)
susceptibilities at zero and non-zero baryon chemical potential.  At
non-zero chemical potential EQCD suffers from a sign problem, albeit
this is milder than in full QCD.  The finite chemical potential
results are obtained by performing simulations with imaginary values
of the chemical potential and then analytically continuing to real
chemical potential.  
We observe that the deviations from the perturbation
theory are significant up to temperatures of order $20\Tc$. On the
other hand, EQCD is observed to work at surprisingly low temperatures:
our results agree well with existing 4d lattice
simulations even slightly below $2\Tc$.  The method also is 
well suited for simulations at non-zero chemical potential, because our 
observations agree those of \cite{Allton:2003vx} and extend 
to even higher values of chemical potential. 
The results have been partly published in
\cite{suskis06,Hietanen:2007ns}.

The paper is organised as follows. In Sec.~\ref{sec:efftheory}
we give the theoretical
background and specify the considered observables. In 
Sec.~\ref{sec:results} we
present the numerical results of lattice Monte Carlo Simulations.
Conclusions are given in Sec.~\ref{sec:conclusions}.

\section{Effective theory}
\label{sec:efftheory}
\subsection{Action}
\label{sec:action}
The electrostatic QCD with finite chemical
potential $\mu$ is defined by the action
\begin{eqnarray}
      \SE & = & \int \dd^3x\LE \nonumber \\
      \LE & = & \frac{1}{2}\Tr[F_{ij}^2]+\Tr[D_i,A_0]^2+m_3^2\Tr[A_0^2]+
      i\gamma_3 \Tr[A_0^3] + \lambda_3(\Tr[A_0^2])^2,
      \label{action}
\end{eqnarray}
where $F_{ij}=\partial_i A_j - \partial_j A_i + ig_3[A_i,A_j]$ and
$D_i=\partial_i+ig_3A_i$. $F_{ij}$, $A_i$ and $A_0$ are traceless
$3\times 3$ Hermitean matrices ($A_0=A_0^aT_a$, etc).  The 
theory has 4 parameters: $g_3^2$ (3-dimensional
gauge coupling), $m_3^2$, $\lambda_3$ and $\gamma_3$, with
dimensions $[g_3^2] = [\lambda_3] = \text{GeV}$,
$[\gamma_3] = \text{GeV}^{3/2}$ and $[m_3^2] = \text{GeV}^2$.
Non-zero value of the parameter $\gamma_3$, caused
by non-zero quark chemical potential, renders the action complex.
Thus, this theory is not free from the sign problem of finite density
QCD.

It is convenient to define three dimensionless ratios
\begin{equation}
  y = \frac{m_3^2}{g_3^4}, ~~~~~
  x = \frac{\lambda_3}{g_3^2}, ~~~~~
  z = \frac{\gamma_3}{g_3^3},
  \label{xyz}
\end{equation}
leaving only $g_3^2$ dimensionful.  Through the dimensional
reduction process (perturbative matching of suitable observables
in EQCD and real QCD), the parameters of EQCD
become functions of physical 4d
parameters: the temperature $T$ and the chemical potential $\mu$ (the
quark masses are set to zero). The parameters are also functions of
the renormalization scale $\Lambda_{\MSb}$ used in the derivation of
the effective theory.
If we denote the number of quark flavors by $\Nf$, for 
$\Nc=3$ the relations are \cite{Kajantie:1997tt,hart00}:
{
\begin{align}
  g_3^2& =  
  \frac{24\pi^2}{33-2\Nf}\frac{T}{\Lb_g/\Lambda_{\MSb}}\left(1-\sum_{i=1}^{\Nf}\frac{1}{9-
  \Nf}\mathcal{D}(\mub_i)x+ \mathcal{O}(x^2)\right) \\
  x & =  \frac{9-\Nf}{33-2\Nf}\frac{1}{\Lb_x/\Lambda_{\MSb}}\left(1-\sum_{i=1}^{\Nf}\frac{1}{9-
  \Nf}\mathcal{D}(\mub_i)x+ \mathcal{O}(x^2) \right) \\
  y &  = 
  \frac{(9-\Nf)(6+\Nf)}{144\pi^2x}\left(1+\sum_{i=1}^{\Nf}\frac{3}{6+\Nf}\mub_i^2\right)+ \nonumber
  \\
   &\quad  \frac{486 -
  33\Nf-11\Nf^2-2\Nf^3}{96\pi^2(9-\Nf)}\left(1+\sum_{i=1}^{\Nf}\frac{3(7+\Nf)(9-2\Nf)}{486-33\Nf-11\Nf^2-2\Nf^3}\mub_i^2\right)+\mathcal{O}(x)
  \\
  z&  =  \sum_{i=1}^{\Nf}\frac{\mub_i}{3\pi}\left(1+\frac{21+3\Nf}{18-2\Nf}x\right) + \mathcal{O}(x^2),
\end{align}}
where $\mub=\mu/(\pi T)$, and, for small $\mub$, $\mathcal{D}(\mub)\approx
-7\zeta(3)\mub^2/2$, and  
\begin{eqnarray}
  \Lb_g&=&4\pi
  T\exp\left(\frac{-3+4\Nf\log4}{66-4\Nf}-\gamma_\textrm{E}\right),\\
  \Lb_x & = & 4\pi
  T\exp\left(\frac{-162+102\Nf-4\Nf^2+(36\Nf-4\Nf^2)\log(4)}{594-75\Nf+\Nf^2}-\gamma_\textrm{E}\right).
\end{eqnarray}
The dimensional reduction scheme is expected to be valid temperatures
down to $\sim$2--4$T_c$ and chemical potential up to $\mu \sim \pi T$ 
or $\mub \sim 1$. For $\Nf=2$ these values correspond to
$x\sim 0.1$ and $z \sim 0.1$.  At higher
temperatures $x$ becomes rapidly smaller. Hence, the higher order
corrections in $x$ in above formulas become in practice very small, and
we ignore 
corrections $\mathcal{O}(x)$ in above expressions.  
We shall further restrict ourselves to 2 massless quarks,  $\Nf=2$:
\begin{equation}
\begin{split}
  g_3^2 & =  g_3^2|_{\mu=0} \\
  x & =  x|_{\mu=0} \\
  y & =  y|_{\mu=0}\left(1+\frac38 \sum_{i=1}^{\Nf}\mub_i^2\right)
  \equiv y_0\left(1+\frac38 \sum_{i=1}^{\Nf}\mub_i^2\right) \\
  z & =  \sum_{i=1}^{\Nf} \frac{\mub_i}{3\pi}.
\end{split}
  \label{xyzmu}
\end{equation}
See \cite{hart00} for more discussion about the effect of this
approximation.

\subsection{Susceptibility}
We define the quark number susceptibility in EQCD as:
\begin{equation}
  \chi_{3,ij}  =   \frac{1}{V}\frac{\partial^2}
  {\partial \mub_i \partial\mub_j}\ln{\cal Z} 
   =   \frac{1}{V}\frac{\partial^2}{\partial \mub_i
    \partial \mub_j}\ln\int \mathcal{D}A_k\mathcal{D}A_0\exp\left(-\SE\right),
  \label{suscdef}
\end{equation}
where $i,j$ stands for quark flavors u and d, and label 3 indicates
that this is a result from 3-dimensional effective theory.
Thus, there are
two independent components of the susceptibility: diagonal ($i=j$) and 
off-diagonal ($i\neq j$). 
Using the shorthand notation for the dimensionless volume averages
\begin{equation}
  \aA^n \equiv \frac1{g_3^n V} \int d^3 x\,\, \Tr A_0^n(x),
  \label{Avolumeavg}
\end{equation}
and defining the condensates
\begin{align}
  C_1 &= \ev{\aA^2} \nonumber \\
  C_2 &= Vg_3^6 \left(\ev{(\aA^3)^2}-\ev{\aA^3}^2 \right) \nonumber \\
  C_3 &= Vg_3^6 \left(\ev{(\aA^2)^2}-\ev{\aA^2}^2 \right)
  \label{condef} \\
  C_4 &= Vg_3^6 \left(\ev{\aA^3\aA^2}-\ev{\aA^3}\ev{\aA^2}\right)\,, \nonumber
\end{align}
we can write the susceptibility as
\begin{equation}
  \frac{\chi_{3,ij}}{g_3^6}  = 
   -\frac{3}{4}\delta_{ij}\,\,y_0\, C_1
   -  \frac{1}{9\pi^2} C_2
   + \frac{9}{16}\mub_i\mub_j\,y_0^2\, C_3
   + i \frac{1}{4\pi}( \mub_i+\mub_j)\,y_0\, C_4
  \label{suskis}
\end{equation}
We note here the rather striking fact that the expectation value
in $C_4$ is {\em purely imaginary\,} for real
$\mub$, rendering the full expression real.  The imaginary expectation
value comes from the complex measure; $\aA^3$ and $\aA^2$
itself are always real-valued.

\subsection{Analytic continuation}
\label{sec:analcont}

The sign problem of finite density QCD is manifested here as an
imaginary term in the EQCD action, \eq{action}.  This makes the
standard Monte Carlo importance sampling impractical, except for
very small chemical potentials and/or small volumes.  
One option to circumvent this problem is to use analytic continuation
to complex values of $\mub$: the sign problem vanishes for purely
imaginary $\mub$.  

However, we emphasize that the direct analytic continuation 
in $\mub$ is clearly suboptimal and unnecessary in this case: 
of the terms appearing in EQCD action \eq{action}, only 
$i\gamma_3 \Tr[A_0^3]$ is responsible for the sign problem.
Thus, it is sufficient to analytically continue $\gamma_3$ (or $z$)
to imaginary values and leave the other parameters to the
values determined by the desired value of $\mub$.  By
far the dominant effect of non-zero $\mub$ is due to the $\mub$-dependence
of the parameter $y$ in \eq{xyzmu}, we can take into account almost
all of the effects of the chemical potential by just using the
correct $y(\mub)$.  The remaining small corrections
are then taken into account by analytic continuation 
$z\rightarrow iz$.%
\footnote{%
  In \cite{suskis06} the 
  susceptibility was evaluated by ignoring this correction; the improved
  statistics here make the small correction non-negligible.}

Because the action (\ref{action}) is invariant under the 
simultaneous change $z \rightarrow -z$ and $A_0\rightarrow -A_0$, 
the partition function must be an even function of $z$ 
(and $\mu$).  From this follows that the expectation values
$\ev{\aA^n}$ are even (odd) functions of $z$ for even (odd) $n$.
Therefore, we can Taylor expand the condensates $C_i$ appearing in the
expression for the susceptibility (\ref{suskis}) in powers of $z$ as
appropriate:
\begin{equation}
  C_i(z) = \sum_n c_{i,n} z^n = \sum_n i^n c_{i,n} (-iz)^n.
  \label{analcont}
\end{equation}
The analytic continuation now proceeds as follows: we perform
simulations with imaginary value of $z$ and determine the
Taylor series coefficients $c_{i,n}$ for each of the condensates
up to the desired order.  Using \eq{analcont} we obtain the 
the condensates $C_i$ at real values of $z$, which 
can be inserted in \eq{suskis} in order to obtain the 
susceptibility.

The dependence of the condensates on $z$ is very mild, as expected, 
and it turns out to be sufficient
to expand the condensates to very low order:
%
%
%
\begin{align}
  C_1 & = a_1 + a_2 z^2   &
  C_3 & = a_4  \nonumber \\
  C_2 & = a_3  &
  C_4 & = 
  \frac{\partial C_1}{\partial(iz)} = -2 i a_2 z .
  \label{cexpansion}
\end{align}
Note that we assume that $C_2$ and $C_3$ are independent
of $z$.  This is indeed the case to the statistical accuracy we
can reach. 

If we now denote with $C_i(\iz)$ the condensates measured from
simulations with imaginary $z=(0,\iz)$, the susceptibility
at real $z=(\rz,0)$ becomes
\begin{equation}
  \label{suscont}
  \begin{split}
    \frac{\chi_{3,ij}(\rz)}{g_3^6} &= 
    -\frac{3}{4}\delta_{ij}\,\,y_0\, 
    \bigg( C_1(\iz) + \frac{\rz^2}{\iz} C_4(\iz) \bigg)
    -  \frac{1}{9\pi^2} C_2(\iz)  \\
    & + \frac{9}{16}\mub_i\mub_j\,y_0^2\, C_3(\iz)
    + \frac{1}{4\pi}( \mub_i+\mub_j)\,y_0\,\frac{\rz}{\iz} C_4(\iz)\,.
  \end{split}
\end{equation}
We note here that one simulation at some $\iz$ is sufficient to obtain
the condensates and the susceptibility at all (small enough) $\rz$.
However, because both $y$ and $z$ depend on $\mub$, only the value of
$\rz$ which corresponds to $\mub$ used in evaluating $y$ is physical.
Thus, for each value of the chemical potential we need to do a new
simulation.  
We also choose to use $\iz = \rz$ in our simulations, eliminating
the ratios $\iz/\rz$ in \eq{suscont}.  In what follows we shall
use the notation $z=\rz=\iz$ to refer to both quantities.

The phase diagram of EQCD has 3 distinct phases: a symmetric phase 
with $\ev{\aA^3} = 0$ and 2 broken phases with non-zero $\ev{\aA^3}$, 
related by reflection 
$\ev{\aA^3} \leftrightarrow -\ev{\aA^3}$ \cite{Kajantie:1998yc}. In order to properly
represent 4d QCD, the effective theory must remain
in the symmetric phase.  In the absence of the chemical potential
the symmetric phase is at most metastable, when the parameters $x$ and $y$ 
are fixed to values which correspond to 4d QCD.  This is normally
not a problem, because the metastability is very strong and for 
all practical purposes the symmetric phase remains stable.  

Applying imaginary chemical potential to the full action would
decrease the value of the parameter $y(\mu)$, \eq{xyzmu}.  Hence, the
metastability would be reduced and finally completely lost at some
value of imaginary $\mu$.  However, for our method of analytic
continuation this problem is completely avoided: because we calculate
$y(\mu)$ with real $\mu$, the value of $y$ increases as $\mu$
increases.  Thus, the physical symmetric phase remains stable
at all values of $\mu$.

\subsection{Relation to 4d physics}
\label{sec:4d}

The relation between $\chi_{3,ij}$ and the physical 4d susceptibility is 
given by
\begin{equation}
  \frac{\chi_{ij}}{T^2}=\frac{g_3^6}{\pi^2T^3}\chi_{3,ij}+\frac{\partial^2}{\partial \mu_i
  \partial \mu_j}\Delta p,
\end{equation}
where $\Delta p=p_{\rm{QCD}}-p_{\rm{EQCD}}$ is the perturbative
3d$\leftrightarrow$4d matching coefficient for pressure.  This is
perturbatively computable order-by-order in coupling constant
expansion, because all perturbatively problematic infrared
singularities of high temperature QCD are fully contained in EQCD.
The matching coefficient is currently known to order
$\mathcal{O}(g^5)$ \cite{gsixg}.%
\footnote{For the pressure the matching coefficient
  has been calculated to $\mathcal{O}(g^6)$ in a much simpler
  theory in Ref.~\cite{Gynther:2007bw}.}

The simulation results in Sec.~\ref{sec:results} indicate that the
$\mathcal{O}(g^6)$ and higher order contributions to the matching
coefficient are very small; indeed, if we compare our results with the
4d simulation results, we obtain an excellent fit when we assume that
these contributions vanish.  Thus, the $\mathcal{O}(g^6)$ and above
contributions to the susceptibility are strongly dominated by the
contributions coming from EQCD.

Because EQCD is derived using perturbation theory, the final results
depend on the perturbative scale $\Lambda_{\MSb}$.  We shall use here
the value $\Lambda_{\MSb} = 245$\,MeV, which has been obtained from
lattice simulations with 2 light Wilson quarks
\cite{DellaMorte:2004bc}.  For the critical temperature we use $\Tc =
170$\,MeV, yielding the ratio $\Tc/\Lambda_{\MSb}=0.7$.\footnote{%
  We obtain the same value by using the results  
  $r_0 T_c = 0.438$ \cite{Bornyakov:2007zu} and 
  $r_0 \Lambda_{\MSb} = 0.62$ \cite{DellaMorte:2004bc}.}
The comparison between EQCD and 4d QCD simulation results is somewhat
sensitive to the precise value of this ratio, but it can vary $\pm
10$\% without significantly affecting the quality of the match.  
The value 0.7 turns out to be close to the optimal one for the matching.

Due to the perturbative nature of the matching equations it turns
out to be convenient to do the matching by subtracting
the known 3d perturbative susceptibility and adding the
4d one:
\begin{equation}
  \frac{\chi}{T^2}=\frac{g_3^6}{\pi^2T^3}\left(\chi_3^{\rm
      latt}-\chi_{3}^{\rm pert}\right)+\frac{\chi^{\rm pert}}{T^2}.
  \label{match}
\end{equation}
Here $\chi_{3}^{\rm pert}$ and $\chi^{\rm pert}$ are 
3d and 4d perturbative results.  
We also note that the quantities 
$\chi_{\rm uu}=\chi_{\rm dd}$ and $\chi_{\rm ud}=\chi_{\rm du}$ are
related to those used in \cite{Allton:2003vx} by
\begin{eqnarray}
  \chi_{\rm q}&=&2(\chi_{\rm uu}+\chi_{\rm ud})\\
  \chi_{\rm I}&=&\frac{1}{2}(\chi_{\rm uu}-\chi_{\rm ud})\\
  \chi_{\rm C} &=&\frac{5}{9}\chi_{\rm uu}-\frac{4}{9}\chi_{\rm ud}.
\end{eqnarray}

\subsection{On the lattice}

The theory in \eq{action} is discretized in a standard way, as 
described in \cite{Kajantie:1997tt}.  Due to the
superrenormalizability of the 3d theory the couplings $\lambda_3$ and
$g_3^2$ do not run,
and $m_3^2$ has well-known linear and logarithmic divergences as 
the lattice spacing $a \rightarrow 0$.  When these divergences
are subtracted the continuum limit is straightforward. 

The evaluation of the quark number susceptibility  requires
the measurement of the condensates in \eq{suskis}
on the lattice. Due to the superrenormalizable nature of
the theory,  measurements can be rigorously converted to $\MSb$ scheme
in the lattice continuum limit; because $\MSb$ was used in in the
perturbative matching to 4d QCD, this also allows us to compare to 4d
results.

The relations between the condensates on the lattice and in continuum
can be written in the limit the lattice spacing $a \rightarrow 0$ (or
$\beta\equiv 6/(g_3^2 a) \rightarrow \infty$) as \cite{Kajantie:1997tt,Laine:1997dy}
\begin{equation}
  \label{latconta}
  \begin{split}
    C_{1,\MSb}  &=  
    C_{1,a} 
    - \tilde{c}_1\beta - \tilde{c}_2\left(\ln\beta +
      \tilde{c_2}'\right) + \mathcal O(1/\beta),  \\
    C_{2,\MSb} &=
    C_{2,a}
    - \left[\bar{c}_2\left(\ln\beta + \bar{c}'_2\right)\right]
    + \mathcal O(1/\beta), \\
    C_{3,\MSb} &=   C_{3,a} + \mathcal O(1/\beta), \\
    C_{4,\MSb} &=   C_{4,a} + \mathcal O(1/\beta).
  \end{split}
\end{equation}
Here labels ${\MSb}$ and $a$ indicate that the quantity
is calculated in $\MSb$ or lattice regularization, respectively.
The numerical coefficients are
\begin{equation}
  \begin{split}
    \tilde{c}_1 &  \approx   0.1684873399, \\
    \tilde{c}_2 & =  
    \frac{3\dA}{(4\pi)^2} \approx  0.1519817755, \\
    \tilde{c}'_2 & \approx  0.66796(1), \\
    \bar{c}_2 & =  \frac{5}{16\pi^2} \approx 0.0316628698900405,\\
    \bar{c}'_2 & \approx  0.08848010.
  \end{split}
\end{equation}

\section{Lattice simulations}
\label{sec:results}

\begin{table}
  \begin{center}
    \begin{tabular}{|l|l|l|}
      \hline
      $T/\Tc$ & $y_0(T)$ & $x(T)$ \\
      \hline
      $ 1.32  $              & 0.357 & 0.13   \\
      $ 2.31  $              & 0.448 & 0.10   \\
      $ 11.5 $              & 0.711 & 0.06   \\
      $ 204  $              & 1.18  & 0.035  \\
      $ 3600 $              & 2.02  & 0.020  \\
      $ 2.4\times 10^7$     & 3.09  & 0.013  \\
      $ 6.2\times 10^9$     & 3.99  & 0.010  \\
      $ 1.9\times 10^{13}$  & 5.31  & 0.0075 \\
      $ 6.1\times 10^{16}$  & 6.62  & 0.006  \\
      \hline
    \end{tabular}
  \end{center}
  \label{tab:temp}
  \caption[a]{The temperatures and corresponding $y_0, x$-values used
    in the simulations.  For each temperature quark chemical potential
    has 6 values, parametrized by
    $z\equiv 2 \mu_q/(3\pi^2 T) =0, 0.025, 0.05,
    0.075, 0.1, 0.15$, and for non-zero $z$ $y$ is modified according
    to \eq{xyzmu}.  At each $(T,\mu)$-pair the simulations
    are done using 6 different lattice spacings, parametrized by
    $\beta \equiv 6/(g_3^2 a) = 32, 40,54, 67, 80, 120$.  
  }
\end{table}

The lattice simulations were carried out using 
two massless quark flavors 
($\Nf=2$). We used nine
different values of temperature $T$, varying from $T\approx 1.9
\Lambda_{\MSb}$ up to $\sim 9\times 10^{16}\Lambda_{\MSb}$.  
The temperature values are shown in Table~\ref{tab:temp}.  While the
largest temperature is huge in physical units, in 3d parameters
the variation is much milder; this is related to the 
fact that QCD approaches weakly coupled theory at high $T$ extremely 
slowly.  Thus, an extreme range of high temperatures is required in
order to reliably assess the convergence to the perturbation theory.

At each temperature we use 6 values for $z = (\mu_{\rm u}+\mu_{\rm d})/(3\pi^2T)$,
$\mu_{\rm u} = \mu_{\rm d}$, up to $z=0.15$ or $\mu_{\rm u}/T \approx 2.22$.  This amounts to
54 different $(T,\mu)$ pairs.  For each physical
point simulations are done using 6 lattice spacings, 
parametrized by $\beta = 6/(g_3^2 a) = 32 \ldots 120$.  Thus,
the lattice spacings vary by almost a factor of 4, enabling
reliable extrapolation of the continuum limit.  For the smallest
lattice spacing ($\beta = 120$), the largest lattice 
size varies between $256^3$ to $320^3$.  

In addition to the simulations at physical parameter values, we also
did a several series of runs at fixed $x$, $y$ and varying $z$.  While
these simulations 
do not correspond to any physical parameter set, they enable us to look
at the $z$-dependence of the condensates separately.
All in all, our dataset contains 693 individual runs.

\subsection{Continuum extrapolation}

It turns out that the accuracy requirement are so high that the
continuum limit extrapolation of the condensates have to be
taken with great care.  Especially
the continuum extrapolation of $\ev{\aA^2}$ is critical, because it
strongly dominates the susceptibility.  While we know the
divergent (as $a\rightarrow 0$) and constant contributions appearing
in the continuum limit, \eq{latconta}, $\mathcal{O}(a)$, $\mathcal{O}(a\ln a)$ and
higher order terms are not yet known.  Thus, we use an ansatz
\begin{equation} 
  \ev{\aA^2}_a - C.T. =
  c_1+\frac{c_2}{\beta} + {\frac{c_2'}{\beta}\log(\beta)} 
  + \frac{c_3}{\beta^2},
\label{logextra}
\end{equation}
where $C.T.$ indicate the known counterterms in \eq{latconta} and
$c_i$ are fit parameters.  
The existence of the
logarithmic term in the ansatz increases the errors of the
extrapolation an order of
magnitude compared to the case without the logarithmic term.
However, $c_2'$ is expected to be a constant independent of $y$: by
dimensional grounds the 
expansion of $\ev{\aA^2}$ in powers of the lattice spacing
can be written as
\begin{equation}
  \ev{\aA^2}_a 
  = \frac{D_1}{a}+D_2g_3^2+ a\left[D_3\,g_3^4 + D_4\,m_3^2 + 
    D_5\,g_3^2\lambda
   + D_6\,\lambda^2\right] + \mathcal{O}(a^2).
\end{equation}
The form of the $\mathcal{O}(a)$ coefficients $D_4$, $D_5$ and $D_6$
is known and they do not contain a term logarithmic in $a$
\cite{Kajantie:1997tt,Laine:1997dy}, whereas the coefficient $D_3$,
which is constant in $y$ (or $m_3^2$), might include one.  Thus,
the possible $a\ln a$ -contribution should indeed be independent
of $y$.  

\begin{figure}
\begin{center}
  \includegraphics*[width=0.5\textwidth]{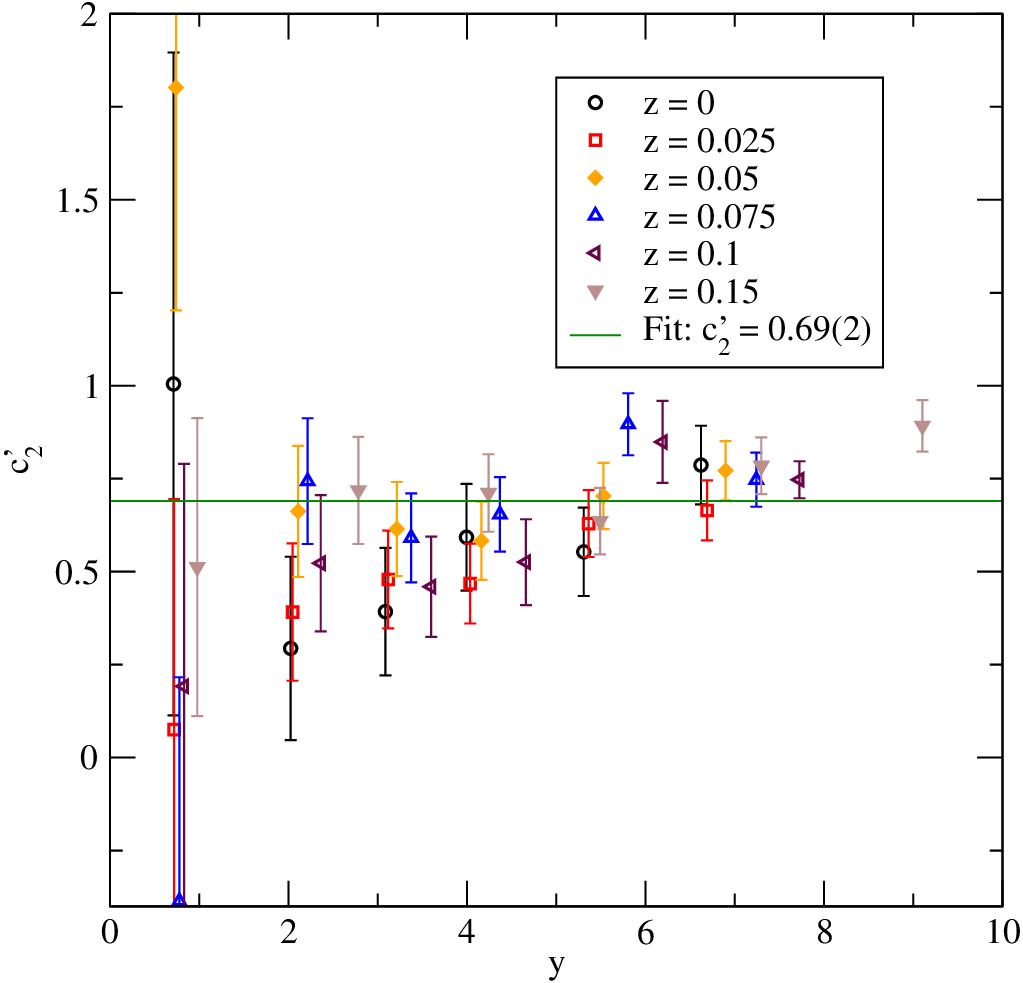}
  \caption{Fitting of the logarithmic coefficient $c_2'$ in continuum
    extrapolation. The data is consistent with the assumption that
    $c_2'$ is a constant $\chi^2/\rm{d.o.f}\approx 57/35$}
  \label{logfit}
\end{center}
\end{figure}

The existence of the logarithmic term can be seen
in Fig.~\ref{logfit}, where we show the parameter $c_2'$ obtained from
continuum fits using the ansatz (\ref{logextra}).  Note that
here $c_2'$ is fitted independently for each physical parameter set,
allowing arbitrary $y$ (and $z$) dependence.
As expected, the result is fairly well consistent with constant 
$c_2' \approx 0.69$; the remaining systematic discrepancies 
in the fit can be caused by contributions
which are of higher order than $\mathcal{O}(a^2)$, including terms of
type $a^2 \ln a$.  Thus, we shall fix $c_2'$ to this value in
\eq{logextra} for all continuum limit extrapolations which follow.

We note that the value of $c_2'$ has negligible effect on the
results at small (physically relevant) temperatures; $c_2'$
could be set to zero without affecting the continuum limit.  
It is significant only at very large $T$, where it potentially 
has a role when we compare simulations with the perturbation theory.
We observe deviations from perturbative results even at very 
high $T$ if $c_2' \lsim 0.4$ (see also \cite{suskis06}).  However,
variations of order $\sim 15$\% around 0.69 do not affect the final
results.  

Nevertheless, it is clear that an analytic calculation
of $\mathcal{O}(a)$ effects in EQCD
would be highly desirable. There is an 
ongoing calculation using stochastic perturbation theory 
\cite{Torrero:2007rs}, which will hopefully confirm our results.  

\begin{figure}
\begin{center}
  \includegraphics*[width=0.99\textwidth]{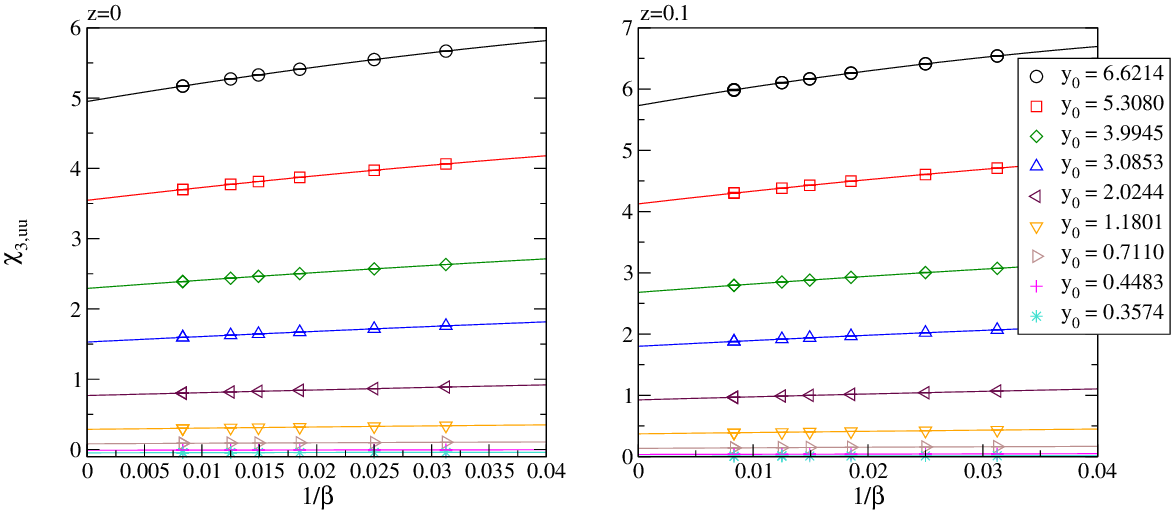}
  \caption{Continuum extrapolation of the
    diagonal susceptibility $\chi_{3,\rm{uu}}$
    at chemical potential $z=0$ and $z=0.1$.
    The statistical errors are too small to be visible.}
\label{fig:extrap}
\end{center}
\end{figure}

The contributions of the other condensates 
are numerically much smaller and
we were not able to see any sign of logarithmic $a$-dependence in 
those.  It turns out that it is advantageous to make the continuum
extrapolation using the full expression of the susceptibility
(\ref{suscont}), instead of extrapolating individual condensates.
(Naturally, after the subtraction of the known counterterms in \eq{latconta}.)
This extrapolation is shown in Fig.~\ref{fig:extrap}.

The lattice volumes are chosen large enough so that finite
volume effects become negligible.  We have tested this by doing 
simulations at selected parameter values using different volumes; 
at the smallest lattice spacing ($\beta=120$) the volume varies 
from $V=144^3$ up to $V=320^3$.   No systematic finite volume effects 
inside two sigma errors. For more discussion of finite size effects
on a related model see Ref.~\cite{plaquette}. 

The finite chemical potential dependence is studied using
the method described
in Sec.~\ref{sec:analcont}.  
The condensates $C_2 = Vg_3^6(\ev{(\aA^3)^2}-\ev{\aA^3}^2)$
and $C_3 = Vg_3^6(\ev{(\aA^2)^2}-\ev{\aA^2}^2)$ should be largely 
independent of $z$ (for fixed $x,y$) for the equation 
(\ref{cexpansion}) to be valid.
This indeed turns out to be the case, within the statistical errors,
and any remnant $z$-dependence is completely drowned out by the
contributions from $z$-independent parts in \eq{cexpansion}.
Indeed, the overall $z$-dependence of each of the condensates in \eq{cexpansion} 
turns out to be statistically almost invisible, with the exception
of  $C_4 = Vg_3^2 (\ev{\aA^2\aA^3}-\ev{\aA^2}\ev{\aA^3})$, which
has a linear $z$-dependence.
In practice the
$\mu$-dependence of the susceptibility is almost completely 
due to the $\mu$-dependence of the parameter $y$ and the
$\mu^2 C_3$-term Eq.~(\ref{suscont})\footnote{This fact was
used in the preliminary results published in ref.~\cite{Hietanen:2007ns}}.  
Nevertheless, here we do take into account the small $z$-dependence 
of $C_1 = \ev{\aA^2}$ and $C_4$, although it will affect the final
results by only about 1 sigma.

\subsection{Diagonal susceptibility}

\begin{figure}
  \begin{center}
    \includegraphics*[width=0.95\textwidth]{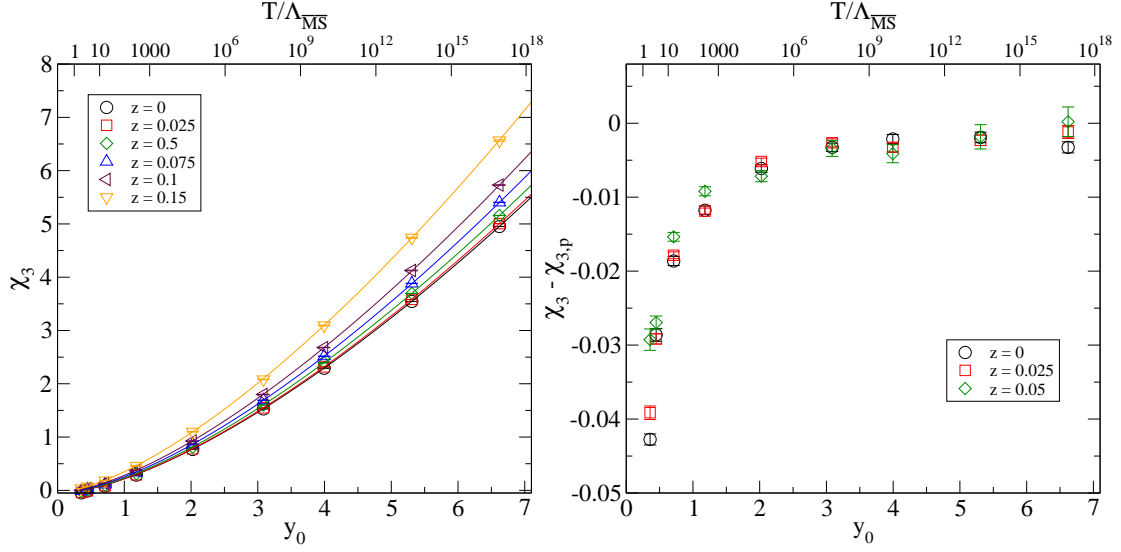}
    \caption{Left: the diagonal quark number susceptibility $\chi_{3,\rm{uu}}/g_3^6$
      at different values of chemical potential.  The symbols indicate
      the lattice
      measurements, and the solid lines are the
      perturbative result.  Right: The difference between
      the lattice and perturbation theory.}
    \label{fig:suskis3d}
  \end{center}
\end{figure}

Now we are in position to compare the continuum limit results with
the perturbation theory.  First we shall look at the diagonal
susceptibility $\chi_{3,\rm{uu}} = \chi_{3,\rm{dd}}$.  The susceptibility has
been calculated in perturbation theory up to order $g^6 \ln 1/g$
 \cite{vuorinen02}.
In 3-dimensional units the perturbative result can be written
as a power series in $1/\sqrt{y_0}$, with the following result:
\begin{align}
  \frac{\chi^{\rm pert}_{3,\rm{uu}}}{g_3^6}&= \frac{8+9\mub^2}{4\sqrt{4+3\mub^2 }} \frac{3y_0^{3/2}}{4\pi}\nonumber \\ 
  & -
  \frac{(9-30x)\mub^2-4(3+10x)+6(4+3\mub^2)\ln(4+3\mub^2)+6(4+3\mub^2)\ln(y_0)}{2(4+3\mub^2)}\frac{3y_0}{(4\pi)^2}
  \nonumber \\
  &  -\frac{(8+3\mub^2)(89+4\pi^2-44\ln(2))}{8(4+3\mub^2)^{3/2}}\frac{9y_0^{1/2}}{(4\pi)^3}
  \nonumber \\
  & 
  +\Big\{576[-3438+40(2+\mub^2)\mub^2] + 119313\pi^2+640(4+3\mub^2)^2\ln(4+3\mub^2) \nonumber \\
  &  \qquad
  +640(4+3\mub^2)^2\ln(y_0)\Big\}\frac{1}{144(4+3\mub^2)^2(4\pi)^4}
  + \frac{80}{3(4\pi)^4} \beta_{\rm M2} + \mathcal{O}(y_0^{-1/2}).
  \label{chidiagpert}
\end{align}
We have set here $\mub_{\rm u}=\mub_{\rm d}=\mub$.  As can be observed in
Fig.~\ref{fig:suskis3d} the overall agreement between the lattice
result and the perturbation theory is very good, especially at large
$y$ (large temperature).  The result contains an unknown
$\mub$-independent order $\mathcal{O}(y_0^0)$ -term denoted by
$\beta_{\rm M2}$ in \cite{vuorinen03}.  The same term appears also in
the off-diagonal susceptibility, \eq{chioffdiagpert}, and it turns out
that it gives much tighter constraints for the value of $\beta_{\rm
  M2}$ than the diagonal one.  The fit to the off-diagonal susceptibility
gives $\beta_{\rm M2}=-0.1 \pm 0.3$.  This value is small enough that
its effect is negligible for the diagonal susceptibility, nonetheless 
we set here $\beta_{\rm M2} = -0.1$.

\begin{figure}[t]
  \begin{center}
     \includegraphics*[width=0.95\textwidth]{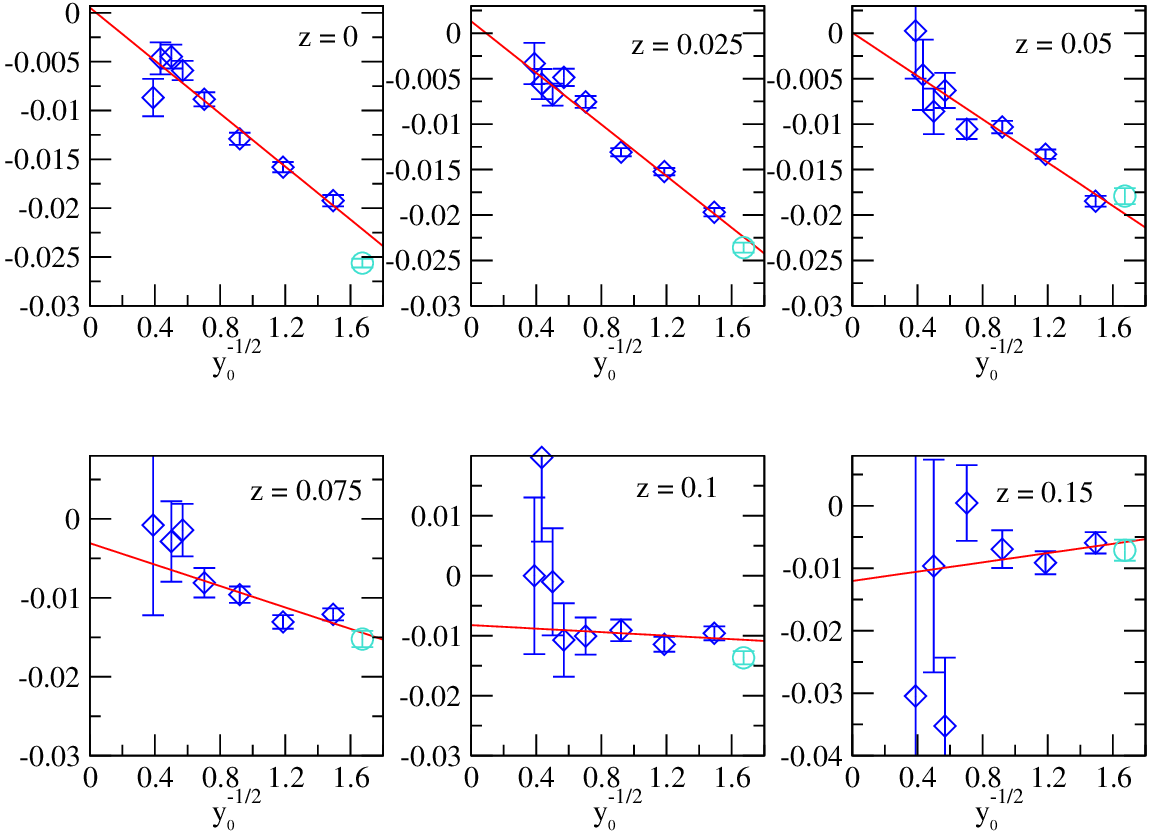}
     \caption{The diagonal susceptibility $(\chi^{\rm lat}_{3,\rm{uu}} -
       \chi^{\rm pert}_{3,\rm{uu}})\sqrt{y}/g_3^6$ as a function of
       $1/\sqrt{y}$ with different values of the chemical potential. Solid
       line is a 1st order polynomial fit.  The data at 
       $y_0^{-1/2}\approx 1.6$
       have been excluded from the fit.
     }
  \label{nppart}
  \end{center}
\end{figure}

In order to quantify the contributions not included in the
perturbative result we calculate the difference $\chi^{\rm latt}_3 -
\chi_3^{\rm pert}$  
and fit a function of form $b_1/y_0^{-1/2} +
b_2/y_0$ to the result.  The fit results are shown in
Fig.~\ref{nppart} and Table~\ref{tnppart}.  
We note that for small
$z$ the $1/y_0^{1/2}$ -term is much smaller than the $1/y_0$-term,
indicating that the $\mathcal{O}(g^7)$ -contribution arising from 
EQCD is smaller in magnitude 
to the $\mathcal{O}(g^8)$ term, at least for all physically relevant
temperatures.  
At large $z$ the statistical errors grow rapidly; this is
due to the term $\propto
z^2 Vg_3^6(\ev{(\aA^2)^2}-\ev{\aA^2}^2)$ in \eq{suscont}.

\begin{table}
  \begin{center}
    \begin{tabular}{|c|c|c|}
        \hline
        z & fit & $\chi^2/$dof \\
        \hline
        0     & $ 0.0008(8)/\sqrt{y_0}-0.0137(8)/y_0$ & 12/6 \\
        0.025 & $ 0.0016(8)/\sqrt{y_0}-0.0143(7)/y_0$ & 18/6\\
        0.05  & $ 0.000(1)/\sqrt{y_0}-0.012(2)/y_0$ & 11/6\\
        0.075 & $-0.003(2)/\sqrt{y_0}-0.007(2)/y_0$ & 17/6\\
        0.1   & $-0.008(4)/\sqrt{y_0}-0.002(3)/y_0$ &  7.5/6\\
        0.15  & $-0.012(6)/\sqrt{y_0}+0.004(5)/y_0$ & 9.7/6 \\
        \hline
      \end{tabular}
      \caption{Fitting a function of form $b_1/\sqrt{y_0} + b_2/y_0$
        to
        $(\chi^{\rm lat}_{3,\rm{uu}}-\chi^{\rm
            pert}_{3,\rm{uu}})/g_3^6$. 
          The smallest $y$ (lowest temperature) points
          are left out of the fit. 
        }
  \label{tnppart}
  \end{center}
\end{table}

\begin{figure}
  \begin{center}
    \includegraphics*[width=0.5\textwidth]{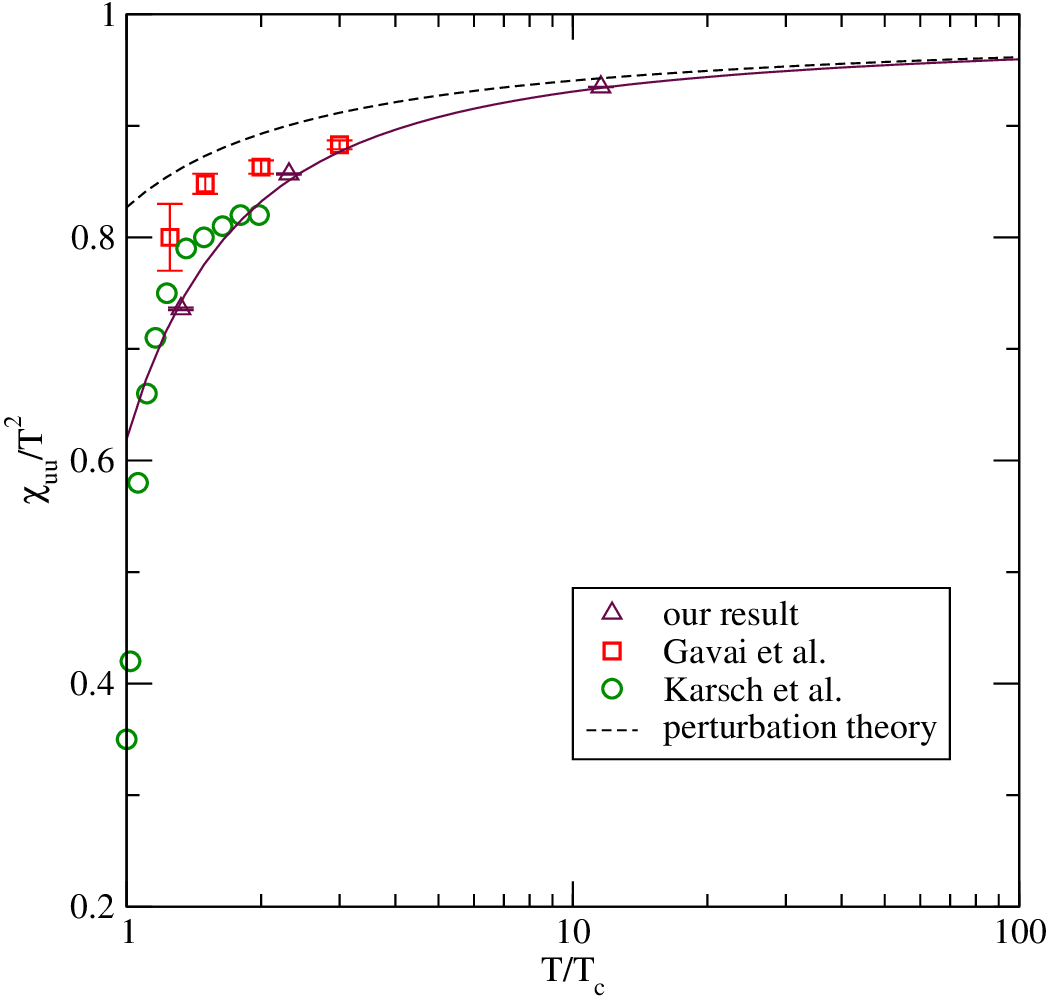}
    \caption{The diagonal susceptibility $\chi_{\rm uu}$ 
      in 4d units at $\mu=0$. The
      data points indicate the full EQCD result.
      The continuous line is the result
      of the fit in Table~\ref{tnppart}.  The dashed line shows
      the perturbative result alone, \eq{chidiagpert}, using the
      same matching as in the EQCD result.  The difference between
      these two curves indicates the magnitude of the non-perturbative
      contributions. The agreement with the 4d-lattice results of
      Gavai et al. \cite{gavai} and Karsch et al. \cite{karsch} is good.}
    \label{suskis4d}     
  \end{center}
\end{figure}

Finally, we obtain the physical 4d result for the diagonal
susceptibility from \eq{match}. 
As described in Sec.~\ref{sec:4d}, the
3d$\leftrightarrow$4d mapping remains sensitive to the unknown
$\mathcal{O}(g^6)$ and higher order perturbative  contributions to the
matching coefficient.  In Fig.~\ref{suskis4d} we show the 
EQCD data at $\mub=0$ with these unknown contributions set to zero.
We observe that the result fits the 4d lattice simulations very well,
clearly indicating that the magnitude of these contributions must
be small, and in what follows we shall set them to zero.
On the other hand,
it should be noted that the difference between the purely perturbative
result and EQCD simulation result is substantial at $T\lsim 10\Tc$,
as indicated by the two lines in Fig.~\ref{suskis4d}.
This is a clear indication that the contributions beyond the 
currently known perturbative ones have non-negligible effect 
at experimentally accessible
temperatures. 

\begin{figure}
  \begin{center}
    \includegraphics*[width=0.45\textwidth]{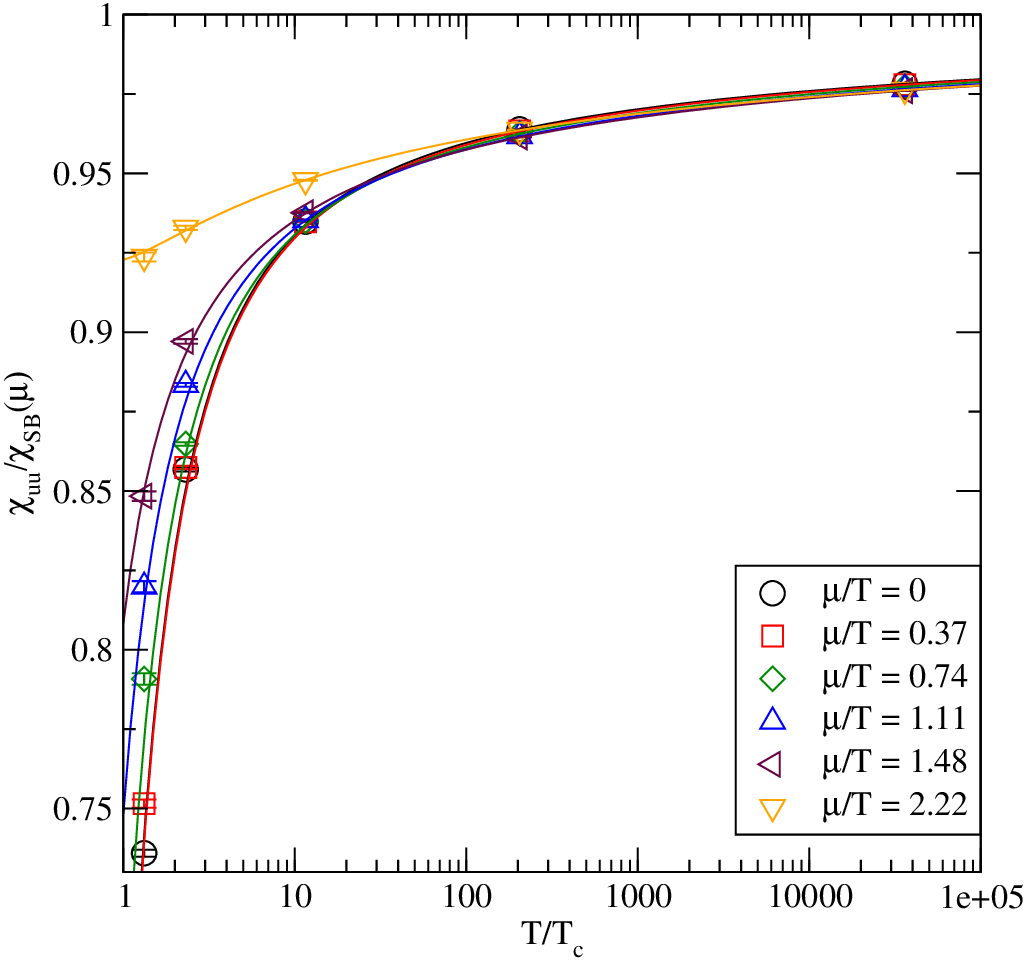}
    \includegraphics*[width=0.45\textwidth]{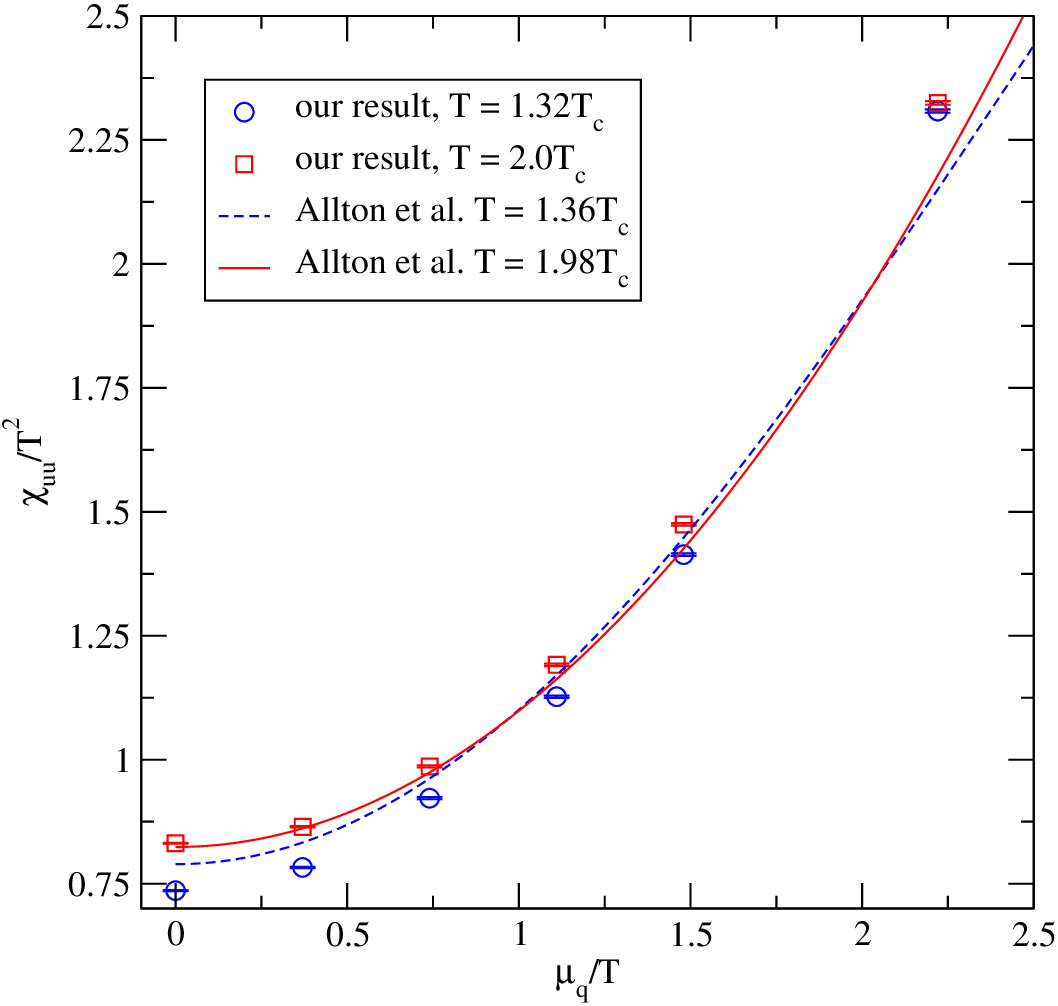}
    \caption{Left: The diagonal susceptibility at different $\mu$,
      normalized to Stefan-Boltzmann law.
      Right: $\mu$-dependence of the susceptibility compared with the 4d lattice results of 
      Allton et al. \cite{Allton:2003vx}.}
    \label{suskis4dmu}
  \end{center}
\end{figure}

The $\mu$-dependence of the diagonal susceptibility is shown
in Fig.~\ref{suskis4dmu}, normalized to the Stefan-Boltzmann value
\begin{equation}
  \chi_{\rm{SB}}(\mu)=T^2+\frac{3}{\pi^2}\mu^2.
\end{equation}
We note that at temperatures above $100\Tc$ the deviation from
the Stefan-Boltzmann law is independent of $\mu$, but at lower $T$ there
is significant $\mu$-dependence.  The $\mu$-dependence matches very
well the 4d lattice results by Allton et al. \cite{Allton:2003vx}, also
shown in Fig.~\ref{suskis4dmu}.

\subsection{Off-diagonal susceptibility}

\begin{figure}
  \begin{center}
    \includegraphics*[width=0.95\textwidth]{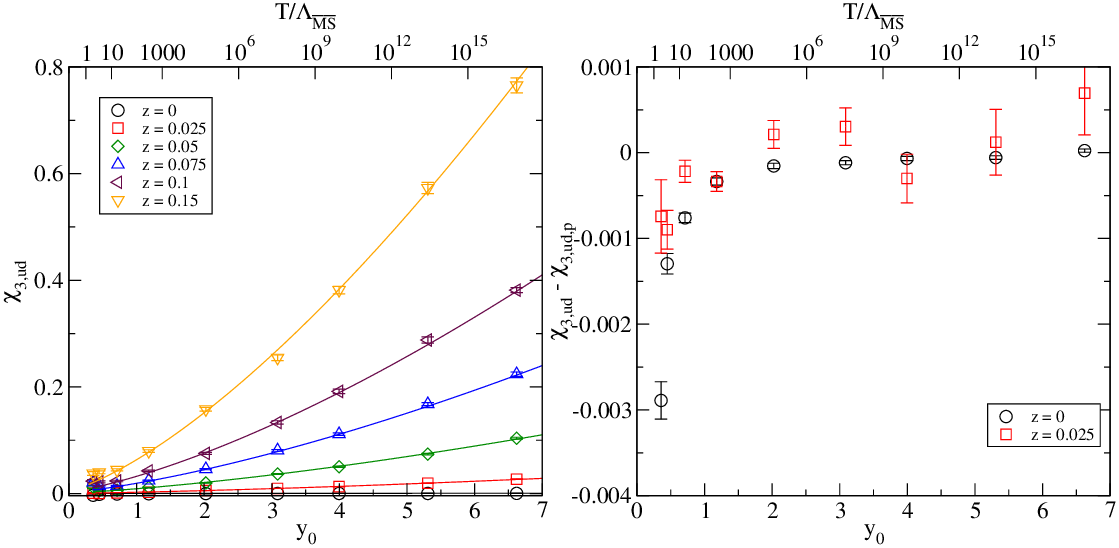}
    \caption{Left: the off-diagonal susceptibility $\chi_{3,\rm{ud}}/g_3^6$ 
      in 3d units.
      Right: the difference between lattice and perturbative susceptibilities
      $(\chi^{\rm latt}_{3,\rm{ud}}-\chi^{\rm pert}_{3,\rm{ud}})/g_3^6$,
      shown at 2 smallest $\mub$.
      The statistical errors grow rapidly as $\mub$ increases. 
    }
    \label{suskisnd3df}
  \end{center}
\end{figure}

The perturbative result for the off-diagonal susceptibility 
in 3d units is
\begin{align}
  \frac{\chi^{\rm pert}_{3,\rm{ud}}}{g_3^2}& =  
  \frac{9\mub^2}{4\sqrt{4+3\mub^2}}\frac{y_0^{3/2}}{4\pi} \nonumber \\
  &  + \frac{27\mub^2}{4+3\mub^2}\frac{y_0}{(4\pi)^2} \nonumber \\
  &  + \frac{27\mub^2(89+4\pi^2-44\log(2))}{8(4+3\mub^2)^{3/2}}\frac{y_0^{1/2}}{(4\pi)^3} \nonumber \\
  &  +
  \Big\{3(2047168-119313\pi^2+15360\mub^2)\mub^2+2560(4+3\mub^2)^2\ln(4+3\mub^2)+\nonumber \\
  & \qquad
  2560(4+3\mub^2)\ln(y_0)\Big\}\frac{1}{576(4+3\mub^2)^2(4\pi)^4}
  + \frac{80}{3(4\pi)^4}\beta_{\rm M2} + \mathcal{O}(y_0^{-1/2})
  \label{chioffdiagpert}
\end{align}
where $\beta_{\rm M2}$ is the same unknown coefficient which
appears in the diagonal susceptibility, \eq{chidiagpert}. 
In this case we can fit the value at $z=0$, obtaining
\begin{equation}
  \beta_{\rm{M2}}=-0.1 \pm 0.3.
\end{equation}
This value is small enough to have in practice negligible effect on the
final results.  Again the simulation data is very well described by the
perturbation theory, Fig.~\ref{suskisnd3df}; only at $z=0$
or at lowest temperatures can we observe deviations from perturbation theory.

\begin{figure}
  \begin{center}
    \includegraphics*[width=0.45\textwidth]{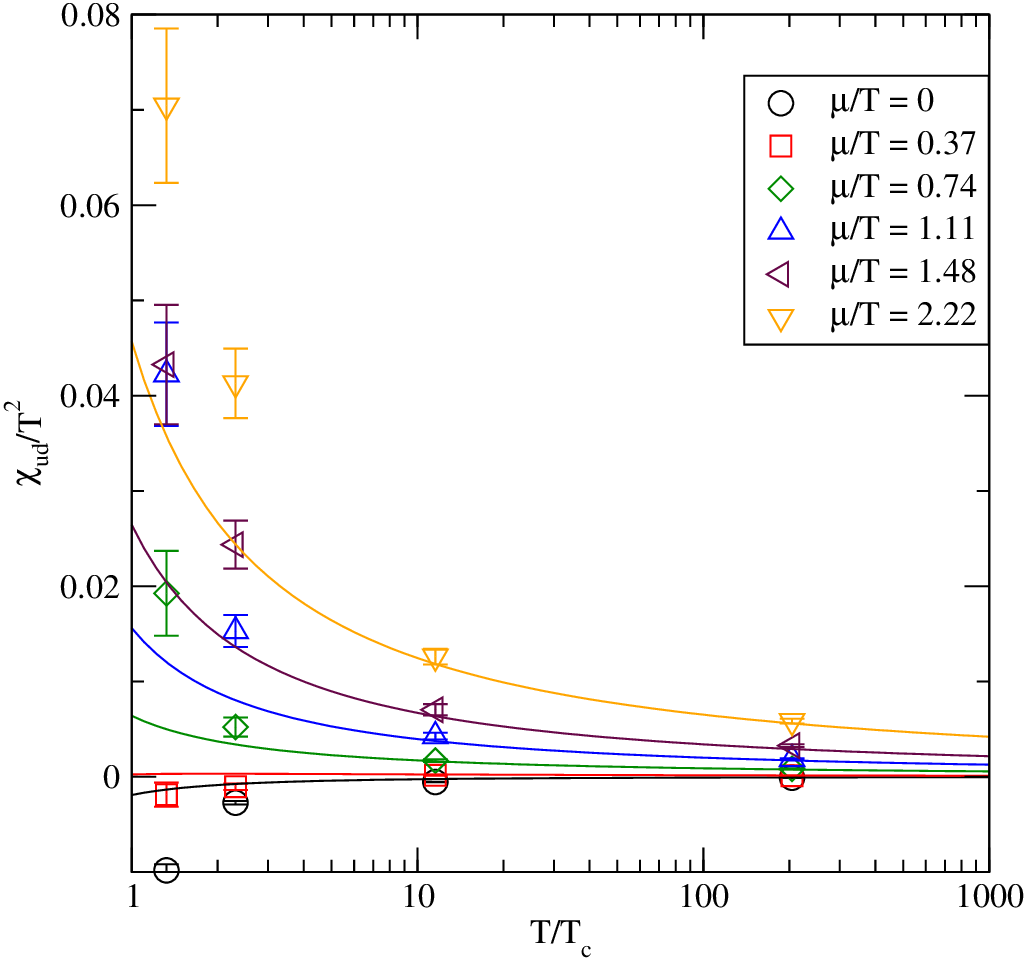}
    \includegraphics*[width=0.45\textwidth]{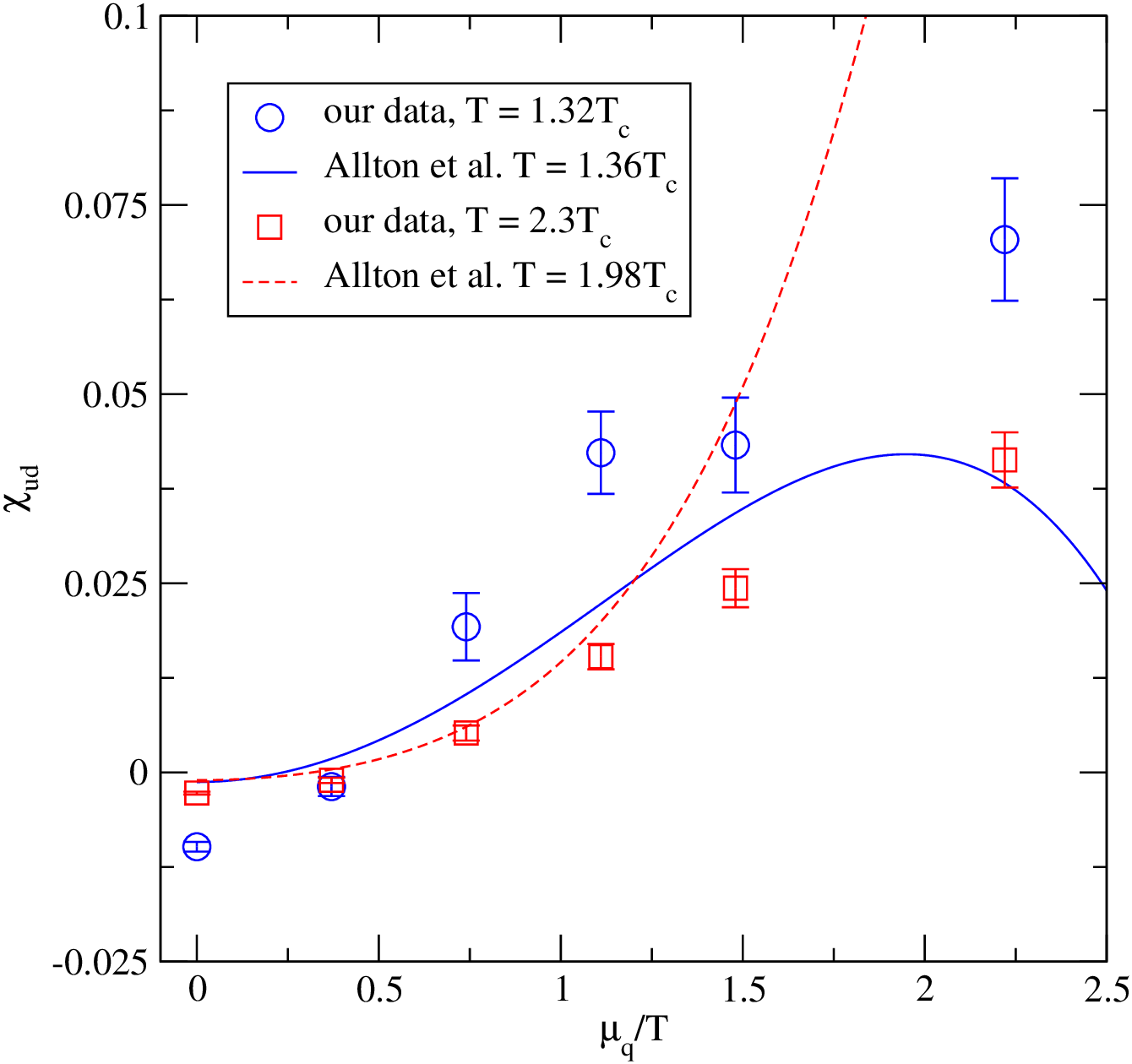}
    \caption{Left: The off-diagonal susceptibility in 4d. At low temperatures
      we obtain significantly different values from the perturbation theory (solid lines),
      but there is no deviation anymore at $T=10\Tc$. Right: $\mu$-dependence of off-diagonal
      susceptibility compared with Allton et
      al. \cite{Allton:2003vx}. The precision of results from
      \cite{Allton:2003vx} are probably not accurate enough to 
      predict the behaviour at region $\mu/T>1$.}
  \label{suskisnd4df}
  \end{center}
\end{figure}

After matching to 4d, we obtain the result for off-diagonal
susceptibility $\chi_{\rm ud}$, shown in Fig.~\ref{suskisnd4df}.
At $T \gsim 10 \Tc$ the results
match the perturbation theory very well, but at lower temperatures
there are deviations: most significantly, at $T=1.32 \Tc$ and $\mu=0$ 
the simulation results
clearly undershoot the perturbation theory. 
On the other hand, the 4d lattice results in \cite{Allton:2003vx} 
at $\mu=0$ indicate 
small but non-zero value, which agrees well with perturbation theory.
\cite{vuorinen02,Blaizot:2001vr}. 
This can be an indication that this point is already outside
the validity range of EQCD; 
however, we also note that by increasing $\Tc/\Lambda_{\MSb}$ the
EQCD results are brought closer to 4d lattice results 
\cite{Allton:2003vx}.  The agreement with the perturbation theory 
and 4d lattice results is rather good already at
$T=2.3\Tc$.  

We also note that the physical value of the
off-diagonal susceptibility
is obtained in EQCD by a subtraction of two divergent as $a\rightarrow 0$
terms; thus, as opposed to full 4d QCD simulation, there is no natural 
approximate symmetry which would force it to be small.  Therefore, 
if EQCD starts to approach the limits of the validity, one can
expect substantial deviations from physical results, as seen 
at $T=1.32 \Tc$.

Nevertheless, the overall $\mu$-dependence of $\chi_{\rm ud}$ is in 
rough accordance with the 4d lattice results \cite{Allton:2003vx}
already at $T=1.32 \Tc$, as shown 
on the right panel in Fig.~\ref{suskisnd4df}, and at $T=2.3 \Tc$
the agreement is already very good.  The non-diagonal susceptibility
is seen to behave quite well up to large values of $\mu/T \sim 2$.

\section{Conclusions}
\label{sec:conclusions}

We have measured the quark number susceptibility of high temperature
finite density QCD using lattice simulations of EQCD, an effective
3-dimensional theory of full 4d QCD.  The very good match to the 4d
lattice results with 2 light quark flavors at low temperatures and
with the perturbation theory at high temperatures shows the wide range
of applicability of the method.  The diagonal susceptibility is seen
to agree with 4d simulations by Allton et al.~\cite{Allton:2003vx}
even below $2 \Tc$, including the dependence on $\mu$.  On the other
hand, we observe a substantial deviation from the known perturbative
result up to temperatures $\sim20\Tc$.  
The off-diagonal susceptibility is compatible
with perturbation theory already at $T \gsim 10\Tc$.  The results also
agree with the 4d simulations \cite{Allton:2003vx} except perhaps at
lowest temperatures, $T< 2\Tc$.

The results clearly indicate that EQCD is a viable method to obtain
quantitatively significant results of the hot QCD plasma down to
$T\sim 2 \Tc$.  Equally significant is the observation that the
currently known perturbative result alone deviates significantly from
the correct result: while the perturbative result can be made to match
the 4d lattice data by adjusting the still unknown (high perturbative
order) matching
coefficients, EQCD allows us to directly measure the differences
between simulations and perturbative calculations without any scale or
matching ambiguities.  Thus, simulations of EQCD are exceptionally
well suited for observing the convergence of the perturbation theory.
It is worth noting that while the EQCD susceptibility also suffers
from matching ambiguity, we obtain an excellent fit to 4d
simulations by assuming these matching coefficients vanish, indicating
that the contribution from these is necessarily very small.

\section{Acknowledgements}
We acknowledge useful discussion with K.~Kajantie, M.~Laine and
A.~Vuorinen. This work has been partly supported by the Magnus Ehrnrooth
Foundation, a Marie Curie Fellowship for Early Stage Researchers
Training, and the Academy of Finland, contract number 114371. 
KR also acknowledges partial support by the 
National Science Foundation under Grant No. PHY05-51164.
Simulations have been carried out at the Finnish IT Center for Science
(CSC).

\end{document}